# Nanoparticle enhanced evaporation of liquids: A case study of silicone oil and water

Wenbin Zhang, Rong Shen, Kunquan Lu, Ailing Ji, and Zexian Cao

## Related Articles

## Additional information on AIP Advances

Journal Homepage: http://aipadvances.aip.org
Journal Information: http://aipadvances.aip.org/about/journal
Top downloads: http://aipadvances.aip.org/most_downloaded
Information for Authors: http://aipadvances.aip.org/authors

# Nanoparticle enhanced evaporation of liquids: A case study of silicone oil and water

Wenbin Zhang, Rong Shen, Kunquan Lu, Ailing Ji, and Zexian Cao[a]
*Institute of Physics, Chinese Academy of Sciences, Beijing 100190, China*

Evaporation is a fundamental physical phenomenon, of which many challenging questions remain unanswered. Enhanced evaporation of liquids in some occasions is of enormous practical significance. Here we report the enhanced evaporation of the nearly permanently stable silicone oil by dispersing with nanopariticles including $CaTiO_3$, anatase and rutile $TiO_2$. An evaporation rate as high as 1.33 mg/h·cm$^2$ was measured in silicone oil when dispersed with 100 nm-sized $CaTiO_3$ particles. Dependence of evaporation rate on the chemistry, size and structure of the particles suggests that some weak absorption sites on the particles half floating on the liquid surface are responsible for the facilitated evaporation of liquid molecules. Enhanced evaporation is also observed for water when dispersed with anatase $TiO_2$ particles. The results can inspire the research of atomistic mechanism for nanoparticle enhanced evaporation and exploration of evaporation control techniques for treatment of oil pollution and restoration of dirty water. *Copyright 2012 Author(s). This article is distributed under a Creative Commons Attribution 3.0 Unported License.* [http://dx.doi.org/10.1063/1.4764294]

## I. INTRODUCTION

Oil pollution has a large devastating impact on the environment. The oil interacts with the surrounding elements via many physical and chemical processes such as oxidation, spreading, evaporation, dissolution, dispersion and biodegradation, etc.[1] These processes proceed in their own characteristic times, but non-exclusively too slow than expected. Provided the rate for anyone of the processes can be considerably promoted, this will then provide a fast and efficient oil removal strategy. Evaporation is one of the most natural processes involved in liquids, and it is also likely to be altered by many factors.[2] If we can find a way to facilitate the evaporation of some non-volatile heavy oils such as silicone oil, this will be of enormous practical importance beyond the treatment of oil pollution. For water, a raised evaporation rate will lead to effective and economic treatment of polluted water, or desalination of seawater.

Evaporation is a surface process. It occurs only on the surface of the liquids and is one of the phenomena classified as interfacial molecular transport, the role of the concrete interfacial conditions during evaporation is still under debate.[3] From the common sense of physics it can be inferred that the evaporation of a tranquil liquid is determined by the surface area, the temperature, and the intermolecular strength for molecules at the outer surface. Raising temperature is the most frequently used strategy for drying in household or in laboratory, but it is impractical in many occasions, and uneconomical.

By observing the evaporation of water from the soil one is easily led to the idea that dispersed particles might enhance the evaporation of water[4]—some soils are particularly draught-prone by loosing water. Liquids in dispersed systems with fine particles may demonstrate a distinct evaporation behavior.[5] The soils of different qualities (composition, dust size, texture, etc.) manifest strikingly dissimilar evaporation behaviors—some are particularly draught-prone whereas those under other

---

[a]To whom correspondence should be addressed. E-mail: zxcao@aphy.iphy.ac.cn

 

conditions can effectively maintain humidity. Enhanced evaporation of water droplets due to surface contamination was noticed many decades ago,[6] but the explanation 'being caused by the ejection of tiny spurts of water, a result of the presence of foreign molecules adsorbed on the surface' is less convincing and far from the possible physical mechanism. It is a notable fact that, despite the importance of evaporation in science and technology, evaporation is poorly understood from a fundamental point of view.[3] The difficulties, in our opinion, lie in the liquid environment where the most modern analyzing tools that penetrate to the level of molecules and atoms fail to work.

What provokes our initial interest on the phenomenon of enhanced evaporation of liquids in a system dispersed with or even dominated by fine particles is the research of electrorheological fluids.[7–9] Electrorheological fluid is a kind of suspension composing dielectric nanoparticles dispersed in non-conducting liquids. By applying an electric field, the electrorheological fluids will immediately undertake a liquid-to-solid-like transition, thus these materials have various potential applications in making shock absorbers, clutches, brakes, artificial joints and so on. Lu *et al.* have fabricated several kinds of high yield stress (even over 200 kPa at 5 kV/mm) electrorheological fluids by dispersing $TiO_2$, $CaTiO_3$ and $SrTiO_3$ nanoparticles in silicone oil.[7,8] However, the implementation of these electrorheological fluids confronts an annoying problem, that is, for some particular combinations of oil and nanoparticle, the oil which is almost permanently stable by itself in the same conditions will disappear very quickly via evaporation.[10] We are thus led to the question as which features of the nanoparticles have caused the fast evaporation of the liquids. The answer to this question is expected to provide clues for the development of strategies and methods for the enhancement or suppress of liquid evaporation, which is of essential importance to, among others, the removal of oil pollution and polluted water, and to ameliorate the quality of draught-prone soils.

In the current work, we report the measurement of evaporation behavior of silicone oil and water when dispersed with $CaTiO_3$, $SiO_2$, anatase and rutile $TiO_2$ nanoparticles of different sizes. Enormously enhanced evaporation can be realized on silicone oil by $CaTiO_3$ and $TiO_2$ nanoparticles, and the effect critically hangs on the nature of particle surface. By meticulously designed experiment we wish to approach the true physical mechanism at microscopic scale as near as possible. The results provide rich clues to the research of atomistic mechanism and to the development of liquid treatment techniques.

## II. EXPERIMENTAL

Suspensions of $CaTiO_3$, $TiO_2$ and $SiO_2$ nanoparticles in silicone oil and in deionized water, were tested. The high-purity silicone oil (Beijing Chemical factory, China) is characterized by a kinetic viscosity of 200 $mm^2$/s, a mass density of 0.97 $g/cm^3$, and a surface tension of 21.1 mN/m at room temperature. Its vapor pressure at 25□ is less than 650 Pa, substantially smaller than that of water (2.81 kPa). Thus the evaporation of this oil under standard ambient temperature and pressure is negligibly slow. $CaTiO_3$ particles, ∼50 nm in size, was self-prepared for the fabrication of high-performance eletrorheological fluids.[11] $TiO_2$ and $SiO_2$ nanoparticles around 50 nm in diameter were purchased from Sigma-Aldrich (USA), and $TiO_2$ nanoparticles of distinct crystal structure, i.e., anatase and rutile phases, of various mean diameters (10, 50, 100 nm) were purchased from MKnano (Canada). The mass densities of the nanoparticles were determined by using the gas displacement density analyzer Accupyc II (Micromeritics Instrument Corporation, Germany). Densities for the nanoparticles here concerned are: $CaTiO_3$, 4.13 $g/cm^3$; $SiO_2$, 2.07 $g/cm^3$; rutile $TiO_2$, 4.23 $g/cm^3$ (100 nm); and anatase $TiO_2$, 3.67 $g/cm^3$ (10 nm), 3.82 $g/cm^3$ (50 nm), and 3.97 $g/cm^3$ (100 nm), respectively. The density of the particles is crucial in determining whether they will precipitate in a liquid.

Glass dish evaporation tests were performed for the various paste-like suspension samples, of various volume fractions which will be specified where they are referred to. The glass dishes (ID40 × 25 mm) containing the samples were kept in a constant climate chamber maintained at 50 °C for all the dispersions in water and the dispersion of $CaTiO_3$ particle in oil, and at 80 °C for the other samples. The exposure area for the samples, calculated from the diameter of the glass dishes, is 12.56 $cm^2$, but it is only a nominal value as the surface of the true dispersion samples can be rumpled. The weight of the samples was recorded for a few days at suitable time intervals by using an electronic balance (Shimadzu, Japan) with a high resolution of $10^{-4}$ g.

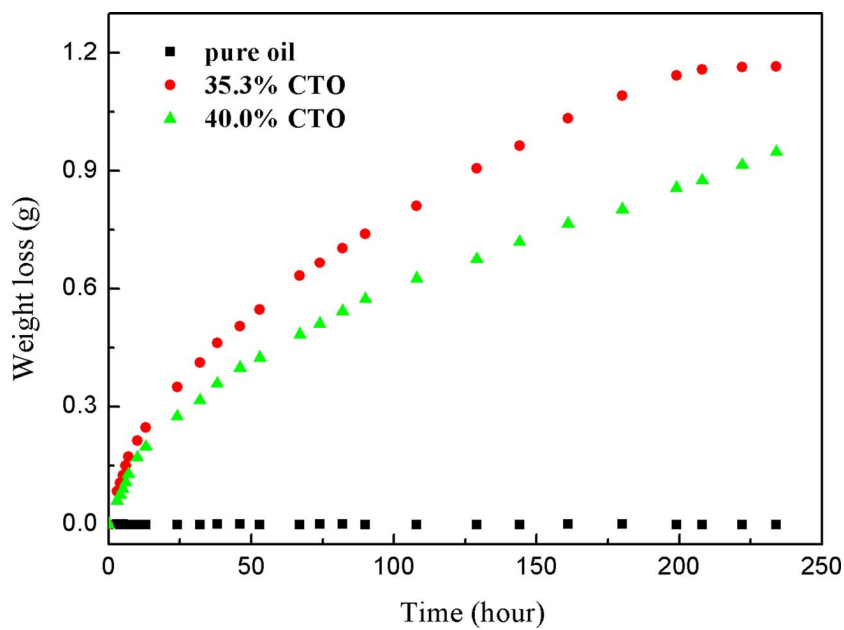

FIG. 1. Weight loss as a function of time for 3 ml (2.91 g) pure silicone oil (square) and for silicone oil plus CaTiO₃ nanoparticle dispersed systems with a particle volume fraction of 35.3% (circle) and 40.0% (triangle), respectively. Environmental temperature: 50°C; humidity: 20%.

## III. RESULTS AND DISCUSSION

Fig. 1 displays the temporal variation of mass loss for three silicone oil samples: one is pure oil (3 ml), and the other two are oil (3 ml) dispersed with 6.75 g and 8.25 g CaTiO₃ nanoparticles—the corresponding volume fractions for the CaTiO₃ nanoparticle are 35.3% and 40.0%, respectively. In the test time as long as 235 h, the mass loss for the pure oil is below the limit of detection, indicating the excellent stability of this material. On the contrary, mass loss due to oil evaporation is remarkably fast for the dispersion samples that for the one with 6.75 g CaTiO₃ nanoparticles, a mass loss of 1.17 g was already measured (Fig. 1, red circle), which amounts to ∼40% of the total mass of oil, 2.93 g. In actuality, the surface of this dispersion sample has become 'dried' at this stage, and the sucking paths for the oil deep inside the sample may be blocke, thus the evaporation comes to a standstill—this is why the curve turns flatted at 210 h. In the first 10 hours, the mass loss rate is 0.021 g/h, this corresponds to an evaporation rate of 1.67 mg/h · cm². But, in interpreting these data we should bear in mind that the paste-like sample has a ruptured surface of which the area can differ a lot from the nominal 12.56 cm².

Remarkably, for samples incorporating CaTiO₃ particles less than 7.0 g, the evaporation of the silicone oil is faster when more CaTiO₃ particles are dispersed in oil. But for the sample of 3 ml oil plus 8.25 g CaTiO₃ nanoparticles (Fig. 1, green triangle), evaporation was slowed down in comparison to the former case that after 235 h, the mass loss is just 0.95 g. In the first 10 h, the evaporation rate is 1.33 mg/h · cm². It seems that at this stage the presence of more nanoparticles leads to the reduction of number of nanoparticles that are involved in the evaporation of liquid molecules.

The results in Fig. 1 undoubtedly indicate that the dispersed nanoparticles induce a very quick evaporation of the non-volatile silicone oil. Such an enormously enhanced evaporation of silicone oil by CaTiO₃ nanoparticles cannot be explained by an enlarged exposure area arising from the floating particles on the liquid surface, as the exposure area can hardly be doubled this way. As the pure silicone oil can be stored for years without noticeable loss, the observation here can be rather more fairly explained that the evaporation of silicone oil is 'provoked' by the presence of CaTiO$_3$ nanoparticles. Something inherent on the oil-particle interface must be responsible for the observed effect.

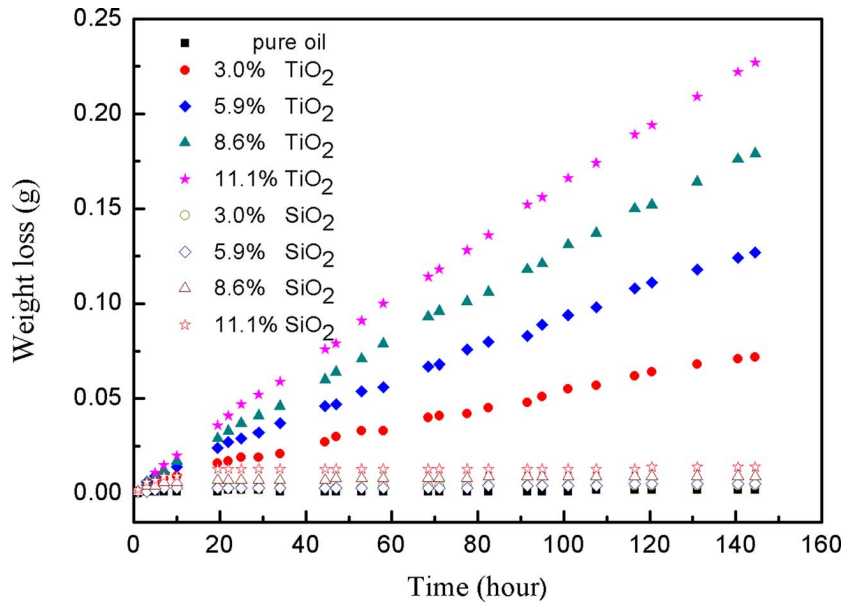

FIG. 2. Weight loss as a function of time for silicone oil (2 ml) dispersed with different amounts of TiO$_2$ and SiO$_2$ nanoparticles. Specified in the graph are the corresponding volume fractions.

To verify this speculation, the same experiment was performed on the dispersed systems of silicone oil (2 ml) plus SiO$_2$ or TiO$_2$ nanoparticles, now the particles have well defined sizes, and the dispersed systems are dilute, having a tranquil, flat surface, thus the evaporation rates can be compared on the same footing (yet the exposure area of 12.56 cm$^2$ is only nominal since the 'liquid surface' of the dispersed system is rugged at the microscopic scale). In the case of SiO$_2$, with the volume fraction of the particles increasing from 3.0% to 11.1% (even larger fraction of particles will cause precipitation), only a negligibly small amount of mass loss, <0.015 g, was measured within 145 h (Fig. 2). Even this small amount of mass loss, observable at larger volume fraction and terminated within 20 h, is most probably ascribed to the loss of water vapor brought into the sample by SiO$_2$ nanoparticles. That means SiO$_2$ nanoparticles have no influence on silicone oil evaporation. In contrast, the effect of TiO$_2$ particles to induce enhanced evaporation of silicone oil is quite obvious, though slower than in the case of CaTiO$_3$. For the sample with a volume fraction of 11.1%, the mass loss at 145 h amounts to ~0.22 g, which is ~23% of the total oil mass. Notably, in the given range of volume fraction, the more particles dispersed in the oil, the faster goes the oil evaporation. All the curves can be well fitted with a straight line, and the evaporation rates of oil are 0.039, 0.068, 0.098, and 0.124 mg/h · cm$^2$ for the volume fraction of nanoparticles of 3%, 5.9%, 8.6%, and 11.1%, respectively. It is to nobody's surprise that in these cases the measured mass loss curves can be well fitted with a line, as the mass loss due evaporation, starting from a zero, will approach a steady-state value via a rising stage characteristic of the mechanism in operation. The dynamic process for the phenomenon concerned here, as for segregation and various other similar situations, can be better reproduced with the function x(t) ∝ [1 − $e^{\alpha t}$erfc($\sqrt{\alpha}$t)] (clearly, it has x (0) = 0), of which the initial part can certainly be fitted by a line.[12]

Evaporation is a surface phenomenon, and if the dispersed nanoparticles can effect an enhanced evaporation, it must be those particles floating on the surface that should claim the credit. Physics intuition tells us that those particles have to be half merging in the liquid and half exposed to the ambient, and the enhanced evaporation occurs on the solid-liquid-ambient boundary where, following the recent model of evaporation,[3] the resistance to the escaping molecules becomes weakened. One possible reason is due to the curved surface morphology, here caused by the floating particles, which has been described by the Kelvin equation. For a curved liquid vapor interface, the equilibrium vapor pressure of the liquid is related to the curvature of the surface via Kelvin equation $\ln \frac{p}{p_0} = \frac{\gamma V_m}{RT}(\frac{1}{R_1} + \frac{1}{R_2})$, where $p_0$ is the equilibrium vapor pressure of a flat surface, $V_M$ is the molar volume, $R_1$, $R_2$ are the principal radii of the curvature, $\gamma$ is the surface tension of the liquid.[13] This equation is found valid with the radius of the liquid drops down to 4 nm. It predicts that undersaturated vapor will condense in channels of sufficiently small dimensions.[14] Our evaporation measurement for silicone oil indicates that its equilibrium vapor pressure has been raised too much, that cannot be explained by a surface with hunchbacks of ∼100 nm in diameter, concave or convex. Moreover, the Kelvin equation describes, in a not very strict sense, a purely geometrical problem. If it is the operating mechanism, the enhanced evaporation in the nanoparticles-dispersed system should not be material dependent, but our observation, to be shown below, tells that this phenomenon even distinguishes between two phases of $TiO_2$ nanoparitcles.

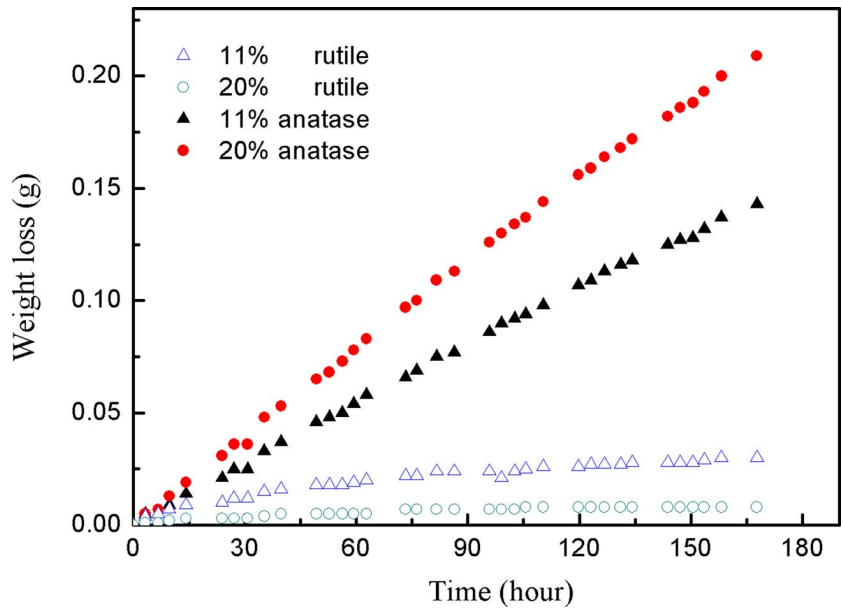

FIG. 3. Comparison of weight losses for the dispersion samples made of silicone oil (2 ml) with anatase and rutile $TiO_2$ nanoparticles. Nominal particle size: 10 nm. Specified in the graph are the corresponding volume fractions.

Assuming some process on the liquid-solid interaction is responsible for the enhanced evaporation of silicone oil, and recalling that the effect is distinct on various solids, it then must hang on some particular properties of the solid surface. For a given crystalline solid, the surface property is first of all phase-dependent. This is anticipated to demonstrate its impact on the enhanced evaporation. To investigate how the crystal structure of $TiO_2$ nanoparticles influences the evaporation of silicone oil, evaporation measurement was carried out on dispersed systems prepared with $TiO_2$ nanoparticles of distinct phases: anatase and rutile. For both phases, the nominal particle size is ∼100 nm, and two samples of a volume fraction of 11% and 20% were tested for each phase. The results are strikingly discernible, as shown in Fig. 3. One sees that the rutile $TiO_2$ particles have only a small effect on the evaporation rate of silicone oil that in 168 h only 0.03 g out of ∼2.0 g oil escaped from the sample of 11% volume fraction, while evaporation from the sample of the 20% volume fraction is simply negligible. On the contrary, the anatase phase $TiO_2$ provokes a greatly enhanced evaporation that the evaporation rate is 0.068 mg/h · $cm^2$ at 11%, and 0.099 mg/h · $cm^2$ at 20%, of the volume fraction.

The anatase and rutile phases of $TiO_2$ are both tetragonal. The anatase structure consists of edge-shared $TiO_6$ octahedra in a tetragonal cell, where as the rutile structure consists of corner-shared $TiO_6$ octahedra in a tetragonal cell. For anatase, a = 3.7845 Å, c = 9.5143 Å, Z = 4, and $\rho$ = 3.9 g/$cm^3$, while for rutile a = 4.5937 Å, c = 2.9587 Å; Z = 2 and $\rho$ = 4.2 g /$cm^3$. The morphology of a crystal particle is determined by its cleavage behavior. Perfect cleavage of anatase occurs on (101) and (001) planes, while for rutile, it shows good cleavage on (110), and moderate cleavage on (001). That the anatase evokes a more effective evaporation of silicone oil may rely on the fact that the (101) plane is the most frequently exposed surface of the anatase TiO2 polymorph,[15, 16] and step edges are the predominant intrinsic defects on surfaces for nanoparticles.[17] It was reported that, for anatase, the adsorption energy of water molecules on (100)-like BI edges are smaller than on the flat (101) surface,[18] and the charge rearrangement at the molecule–anatase interface affects the adsorption of further liquid molecules. The weak absorption of liquid molecules on some sites on the anatase surface, noticing that cleavage occurs along planes of relative weakness, may partially be the reason of enhanced oil evaporation This is consistent with the observation that the step edges on the (101) plane of anatase are also responsible for its photocatalytic ability,[17] and CaTiO3, causing even more serious evaporation of silicone oil, has favorably TiO2-terminated (001) facets. The surface of rutile TiO2 can also provide some weak absorption sites as it also effect the evaporation of silicone oil (see Fig. 3) and the dissociation of ethanol.[19] This speculation is reasonable, yet needs experimental verification on the molecular level. However, due to the small size of the nanoparticles, the liquid environment and the rugged surface (at nanometer scale) of the dispersed system, it will be a challenge to find proper analyzing tool or tool sets to offer experimental verification.

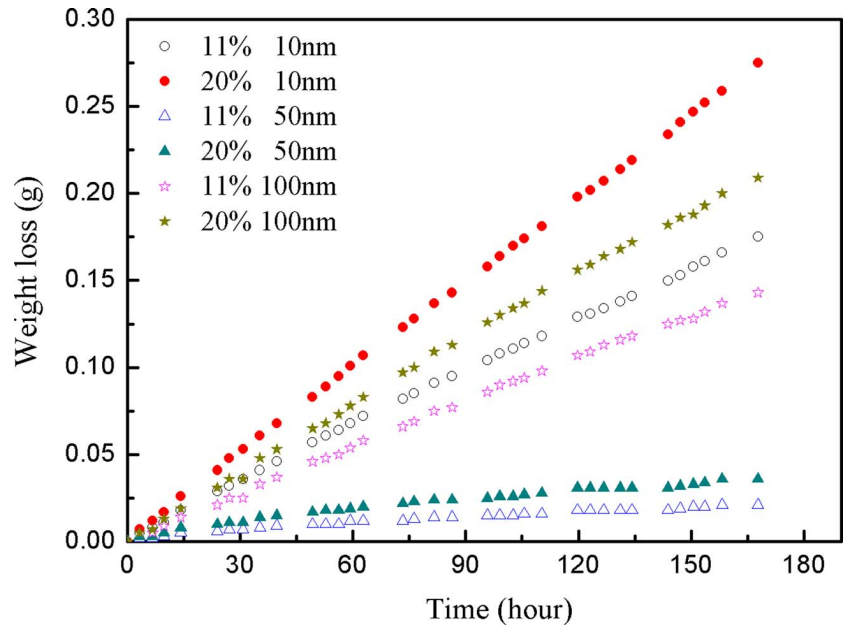

FIG. 4. Comparison of weight loss for the dispersed samples made of silicone oil (2 ml) with anatase TiO2 nanoparticles of various sizes. Particle size and the corresponding volume fraction are specified in the graph.

As the facet configuration of a nanoparticle also hangs on the particle size, it is anticipated that the enhanced evaporation of the dispersion systems may vary with the particle size. We further studied the oil evaporation enhancement of anatase TiO2 nanoparticles in different diameters: 10 nm, 50 nm and 100 nm. For all the three particle sizes, the larger volume fraction (20%) implies a faster evaporation (Fig. 4). For both volume fractions, as a surprise or not, the 10 nm TiO2 nanoparticles have the most pronounced effect followed by 100 nm nanoparticles. The 50 nm nanoparticles are least effective in promoting the evaporation. For the 10 nm particle in a volume fraction of 20%, the evaporation rate amounts to 0.13 mg/h · cm$^2$, whereas for the 50 nm particle the rate becomes down to ~13% of the former. This amazing result indicates the relation of evaporation to the fine microstructure of the liquid-particle interface, which serves a support to our speculation above.

Following the recent model of evaporation where detachment of atoms from the surface occurs when an inside outgoing atom reaches the plane of the surface,[3] the evaporation rate is determined by the intermolecular force and the configuration of molecules at surface which constitutes the resistance to a molecule to break away from surface. For simplicity, we consider the case that the escaping molecules come from the outermost layer and need only overcome the bonding energy to the low-lying neighbors, as shown in the upper left of Fig. 5. It is a simple fact that only nanoparticles floating on the liquid surface could have the chance to enhance the evaporation, and events to enhance evaporation could only occur at the top contact area between nanoparticle and liquid that exposes to the ambient. At this point, it can be conceived that nanoparticles' surface is only partially covered by one or several layers of liquid molecules, and these liquid molecules are adsorbed on some sites of the nanoparticles' surface, as illustrated in the upper right of Fig. 5. If liquid molecules have enough kinetic energy to overcome the adsorption energy $E_{ad}$ at these sites, they will evaporate. Anatase and $CaTiO_3$ can provide some sites of low absorption energy. That $CaTiO_3$ is more effective than anatase can be explained that $CaO_2$-related facets can also be active in providing weak adsorption sites, as the same phenomenon is also observed in the silicone oil plus Ca $(COO)_2$ nanoparticle system (not shown here). For long molecules such as silicone oil (group of -$[Si(CH_3)_2O]_n$), the molecules may partially attached to the solid surface while part of them remains bonded to its comrades in liquid. And reduction of molecule number constituting the resistance to the escaping molecule also results in promoted evaporation.

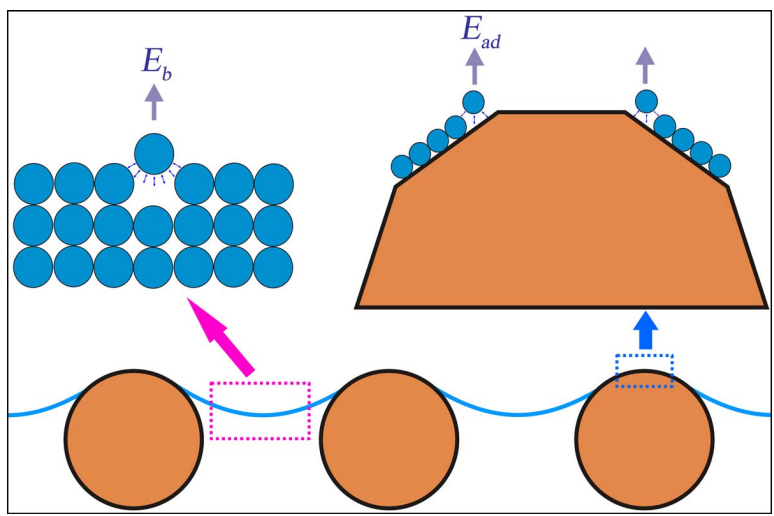

FIG. 5. Sketch of nanoparticles partially immersed in the surface of a liquid, where the curved delineation represents the liquid menisci formed between nanoparticles at the surface (in some cases the most part of the particles may be floating over the liquid surface). Enlarged parts illustrate the distinct environment for the outermost liquid molecules to evaporate. $E_b$: binding energy of liquid molecule at liquid surface; $E_{ad}$: adsorption energy of liquid molecule on nanoparticle.

The discussion above also applies to the evaporation of water. Water in ambient condition evaporates quite quickly, yet enhanced evaporation of water is of importance in the cases of, say, treatment of polluted water. What is quantitative evaporation behavior of water dispersed with some well-defined nanoparticles? We tested the water plus $CaTiO_3$, $SiO_2$, anatase $TiO_2$ or rutile $TiO_2$ particles. As water is much less viscous than silicone oil, the $CaTiO_3$ and $SiO_2$ nanoparticles precipitate in water, thus the effect on the evaporation of water can not be approached in this way. The same precipitation problem occurs to the rutile $TiO_2$ particles if more particles are incorporated. The anatase $TiO_2$ nanoparticles demonstrate a noticeable modification to the evaporation of water. From Fig. 6 one sees that water dispersed with anatase particles (25 nm) does display different evaporation rates. With a volume fraction of particle of 3.0% the evaporation rate comes down to 64% of that for pure water, i.e., 0.23 mg/h·cm² under the current condition, at 5.9% the evaporation rate recovers 90% of the original value. However, with the volume fraction of 8.6%, the evaporation rate rises by a factor of 1.6. Similar effect was observed in the water plus flour system. The enhancement effect to water evaporation by anatase particles, when compared with that of $CaTiO_3$ on silicone oil, is only marginal. The reason for the amazing observation here can be ascribed to the fact that water molecules are polar and small molecules, and as one of the many anomalies for water, the surface is denser than in bulk. The results, while limited, verify the possibility of applying proper nanoparticles to modify the evaporation of water in two directions, which may provide clue to understanding water evaporation from soils of distinct qualities.

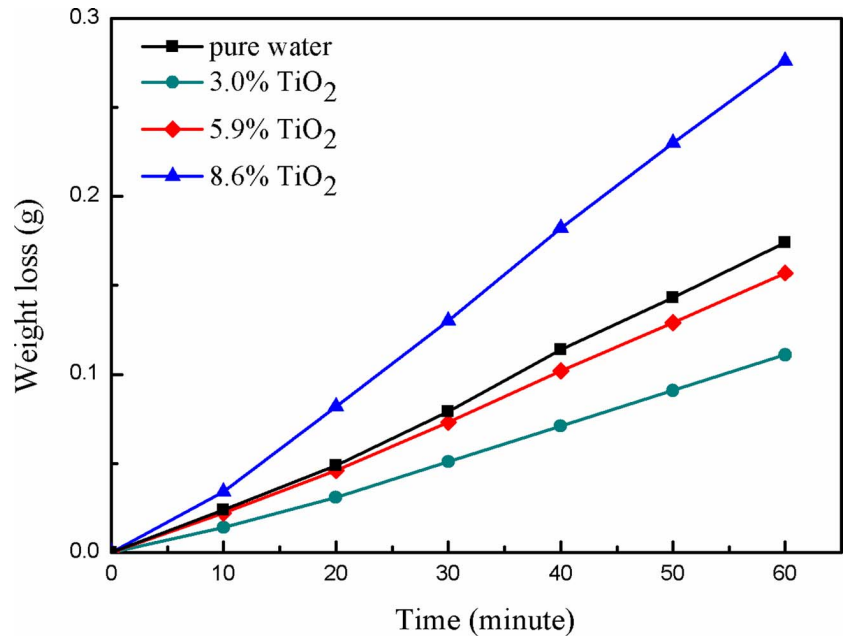

FIG. 6.  Weight loss as a function of time for pure deionized water (2 ml) and water (2 ml) dispersed with different amounts of anatase $TiO_2$ nanoparticles of 25 nm in size. Specified in the graph is the corresponding volume fraction. Environmental temperature: 50 °C; humidity: 20%.

## IV.  CONCLUSIONS

In summary, significantly enhanced evaporation was observed for silicone oil dispersed with $CaTiO_3$ and $TiO_2$ nanoparticles. Our measurement clearly demonstrated that only nanoparticles that half float on the liquid surface and provide some weak adsorption sites, thus depending on the specific nature of the nanparticles's surface, lead to enhanced evaporation, as verified by the variation of evaporation rate on the size and volume fraction of particles, and also differentiated by the anatase and rutile $TiO_2$ phases. The anatase $TiO_2$ particles can both suppress and enhance the evaporation of water depend on the volume fraction. Based on the analysis of the experiment results, one point can be concluded that liquid molecules at the boundary region between the dried and wetted region, when migrate to the weak adsorption site, will have a greater chance to escape via evaporation. The exact atomistic mechanism needs further delicate investigation—a challenging topic considering the fact that the object of concern is partially wetted nanoparticles on liquid surface. The current report aims to bring attention to this interesting and useful phenomenon. These preliminary results give us clues to the selection of material and design of structure for the evaporation control of paste-like dispersed systems, they verify the feasibility of enhanced liquid evaporation by adding properly chosen nanoparticles, which might provide new technologies for treatment of oil pollution and polluted water.

## ACKNOWLEDGMENTS

This work was financially supported by the National Basic Research Program of China grant nos. 2009CB930801 and 2012CB933002, by the Natural Science Foundation of China Grant nos. 10974227, 51172272, and 10904165, and by the Chinese Academy of Sciences.